# Nanoscale
# Advances

Accepted Manuscript

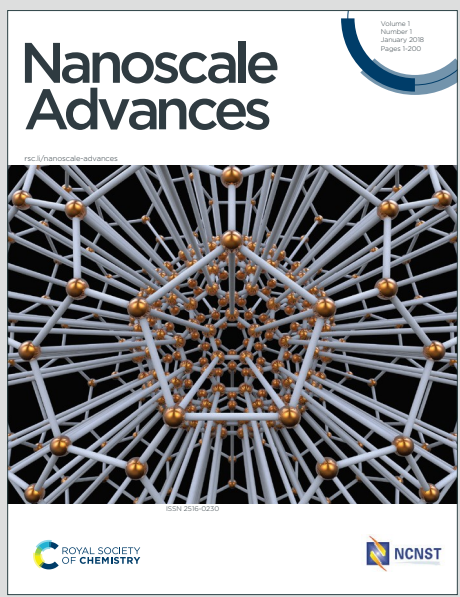

Nanoscale
Advances

rsc.li/nanoscale-advances

ROYAL SOCIETY
OF CHEMISTRY

rsc.li/nanoscale-advances

# ARTICLE

## Nanoscale structure detection and monitoring of tumour growth with optical coherence tomography

Nandan Das*[ad], Alexandrov Sergey[a], Yi Zhou[a], Katie E. Gilligan[b], Róisín M Dwyer[b], Martin Leahy[ac]

Approximately 90% of cancers have their origins in epithelial tissues and this leads to epithelial thickening, but the ultrastructural changes and underlying architecture is less well known. Depth resolved label free visualization of nanoscale tissue morphology is required to reveal the extent and distribution of ultrastructural changes in underlying tissue, but is difficult to achieve with existing imaging modalities. We developed a nanosensitive optical coherence tomography (nsOCT) approach to provide such imaging based on dominant axial structure with a few nanometre detection accuracy. nsOCT maps the distribution of axial structural sizes an order of magnitude smaller than the axial resolution of the system. We validated nsOCT methodology by detecting synthetic axial structure via numerical simulations. Subsequently, we validated the nsOCT technique experimentally by detecting known structures from a commercially fabricated sample. nsOCT reveals scaling with *different depth of dominant submicron structural changes* associated with carcinoma which may inform the origins of the disease, its progression and improve diagnosis.

## Introduction

Many early disease processes in living tissues exhibit changes at the nanometre scale. Non-invasive, label-free detection of submicron structural alterations with nanoscale accuracy in biological tissue poses significant challenges to researchers and healthcare professionals. Therefore, it is highly desirable to be able to detect these nanoscopic changes in submicron structure during physiological processes in real time. Most of the nanoscopic techniques rely on labelling and is limited to superficial imaging which is generally not suitable for *in vivo* applications[1–6]. In the fight against cancer, nanoscopy and nanotechnology can provide molecular level contrast to detect disease earlier and more accurately[7–13]. At the same time, it can help us to monitor the patient improvement under treatment. In recent development, partial wave spectroscopy based approach demonstrated that it can detect nano-sensitive structure and its alteration in cells[14]. Although, this method only can detect overall averaged structure, but not a structure hiding inside scattering tissue depth. There is also some progress in development of true colour depth resolved imaging which is based on absorption contrast of oxy and de-oxy haemoglobin and provide molecular contrast present there rather submicron structural morphology[15]. There is another recent work claiming depth resolved nanoscale mapping for cancer detection[16] which we believe based on overall average length scale present through the depth rather than depth resolved sub-voxel morphological scales. In an early development, optical coherence tomography demonstrated nano-sensitive structural alteration in biological tissue which again based on statistical mass distribution throughout the accessible depth[17]. These techniques only can detect overall nanoscopic changes rather than depth resolved sub-voxel nanoscopic information which is crucial to visualise subtle alteration of local submicron structure for better diagnosis. In this direction, our research group recently developed nano-sensitive optical coherence tomography (nsOCT) techniques which can detect depth resolved submicron dominant structure with nanometre scale accuracy[18–20]. Recently, we have demonstrated nsOCT-based correlation mapping imaging to visualize few nanometre structural changes without any labelling[21] and application of the nsOCT for cornea cross-linking study[22]. Another research group recently applied nsOCT for *in vivo* detection of nanometre-scale structural changes of the human tympanic membrane in otitis media[23]. Theoretical approach of nsOCT can be found in our previous published paper[19]. In this present research, we have numerically validated nsOCT approach on synthetic submicron axial structure with few nanometre accuracies. Results of numerical simulation have been confirmed via experimental nsOCT imaging of two fabricated samples (repeating axial structure 0.43156 µm and 0.44167 µm respectively) and by detecting structure with ~ 5 nm accuracy. Finally, we have applied our developed nsOCT method on tumour and healthy tissue from mouse and found interesting change

*a. Tissue Optics and Microcirculation Imaging (TOMI), National University of Ireland, Galway, Ireland.*
*b. Discipline of Surgery, Lambe Institute for Translational Research, National University of Ireland Galway, Ireland.*
*c. Institute of Photonic Sciences (ICFO), Barcelona, Spain.*
*d. Department of Biomedical Engineering (IMT), Linköping University, Sweden.*
*Corresponding author: nandankds@gmail.com.*

in depth resolved dominant submicron structures as tumour progress. Initial experimental results on mouse mammary fat pad (MFP) demonstrate new possibilities for the label free nano-sensitive detection of sub-voxel ultra-structure corresponding to tumour progress. Thus, nsOCT offers exciting depth resolved optical nanoscopy and far-reaching opportunities for early disease diagnosis and treatment response monitoring.

## Materials and method

### Periodic submicron structure

We have received two experimental samples from OptiGrate Corp. USA, with different axial periodic structure, where the corresponding periods are 431.56 nm and 441.67 nm (**Fig. 1(c)**). Each periodic layer fabricated with sinusoidal refractive index varies 1.483 ± 0.001. No variation in periodicity vs time if temperature of usage will be <400°C (periods are stable better than 1 pm). Thickness of each sample is ~ 3 mm. We have also synthesized three axial periodic structure with 424.8 nm 431.56 nm and 442.67 nm for numerical study (**Fig. 1(a)**). In axial periodic structure synthetization, we have considered same sinusoidal refractive index variation to mimic experimental sample. In additional numerical study, we have synthesized submicron structure with 444.70 nm, 451.90 nm, 462.77 nm, 473.61 nm and 488.10 nm (**Fig. 2(a)** and **Fig. 2(c)**). Here, we have considered sinusoidal refractive index variation 1.383 ± 0.001 to mimic tissue like refractive index variation. In an advance simulation, we have synthesized randomized submicron structure with different depths to mimic tissue (**Fig. 2(e)** and **Fig. 2(f)**). Here again structure varies from 444 nm to 488 nm, and we have considered sinusoidal refractive index variation 1.383 ± 0.001 over each submicron structure.

### Tissue samples preparation

All animal procedures were performed in accordance with the Guidelines for Care and Use of Laboratory Animals of "the Animal Care Research Ethics Committee (ACREC), National University of Ireland Galway (NUIG)" and approved by the "Health Product Regularity Authority (HPRA), Ireland".

The acquisition of the tissue samples does not affect any ethical, health, or privacy concerns and was performed according to the ethical regulations and safety standards of the NUI Galway.Female BALB/c mice (Charles River Laboratories Ltd.) aged between 6 and 8 weeks were employed. Mice received a mammary fat pad (MFP, 4th inguinal) injection of $1 \times 10^5$ 4T1 breast cancer cells suspended in 100µl RPMI medium. The tumours were detected by palpation after seven days of injection and were visually inspected. We believed that grown tumour after seven days still in early stage of cancer. Tumour growth was monitored using callipers measurement, and at the appropriate time animals were sacrificed by $CO_2$ inhalation. The tumour size was 715 mm³. Tumour tissue and healthy portion was harvested in PBS solution separately to take it at nsOCT imaging facilities for *ex-vivo* study. Sample was taken out from PBS and mounted on a glass slide to bring the sample under the objective of spectral domain OCT system (Telesto III, Thorlabs Inc.) for nsOCT acquisition.

## Analysis and detection methodology of nano-sensitive optical coherence tomography (nsOCT)

We and other research group have previously demonstrated a method, Fourier domain nano-sensitive optical coherence tomography (nsOCT), for detection of depth resolved submicron dominant structure non-invasively with nanoscale accuracy[18,19,21]. Our approach is based on finding maximum contributed spatial period (equivalent to size of dominant axial structure), calculated from corresponding OCT signal, at all accessible depth (voxels). This spatial period carries information about dominant submicron (0.444 µm - 0.488 µm corresponding to the wavelength range available in our OCT system) structure within each voxel which are extremely sensitive to structural alterations. Here, we have described the principles of this approach and demonstrates its feasibility by clearly differentiating different submicron (0.444 µm - 0.488 µm) structure. Initially, we have validated our approach by numerically detecting three synthetic submicron structure with ~ 1.5 nm accuracy (see **Fig. 1(a)** and **1(b)**).

For spectral interference data acquisition, the Telesto Ⅲ (Thorlabs Inc.), a commercial spectral-domain optical coherence tomography (sdOCT) system utilized with an objective lens LSM03 (NA = 0.055, lateral resolution ≈ 13 µm). Schematic of the OCT system can be found in electronic supplementary information (ESI: **Fig. S1**). This sdOCT system operating with wavelength range from λ = 1176 nm-1413 nm with centre wavelength $\lambda_c$ = 1300 nm and sensitivity 96 dB @76kHz rate), can achieve an axial resolution of 5.5 µm and imaging depth up to 3.6 mm in air. Detectable spatial period (HZ = λ/2n; n = refractive index of the medium) depends on wavelength range of the broadband source[19]. Therefore, in case of biological tissue, theoretically we can detect 420 nm to 504 nm structure with few nanometre uncertainty depending upon spectrometer resolution[19]. Note that, the range of detectable axial dominant structure depends on wavelength range of OCT system[19]. The possible detection error also depends on how closely we can scan those windows over the spectral range[19]. Detailed flowchart of nsOCT analysis displayed in electronic supplementary information (ESI: **Fig. S2**) and theoretical discussion can be found in our previous publication[19].

## Numerical simulations to construct nsOCT from synthesized submicron structures

We have followed standard FD-OCT theory and simulation strategy to implement proposed nsOCT simulation[24,25]. We have used Matlab 2019a (MathWorks®) for its implementation. In validation purpose, we performed a numerical simulation for periodicity with 424.8 nm in addition to 431.56 nm and 441.67 nm. Each periodic layer fabricated with sinusoidal refractive index varies 1.483 ± 0.001. Each layer divided into many sub layers with thickness 0.01 nm to make it smooth enough in sinusoidal index variation to mimic fabricated layer sample from OptiGrate Corp. USA. For example, if periodicity is 424.8 nm, the number of sublayers in each periodic layer is 42480. The OCT signal was formed as an interference of the reflected light from layer sample with the reference wave from gold mirror. We have also introduced suitable noise in source and detector spectra to mimic experimental reality in our simulation. The signal to noise ratio, SNR ≈ 86 dB in our simulation. The inverse Fourier transform was used to form a B-scan of these samples, presented in **Fig. 1(a)**. For OCT simulation, similar sinusoidal index variation is considered for two different layers of thickness 441.67 nm and 431.56 nm. In addition, we have simulated for 424.8 nm periodic and repeating structure (**Fig. 1(a)** and **Fig. 1(b)**) to demonstrate our state of art methodology for depth resolved submicron dominant structural detection with few nanometre accuracies.

To check abilities of the nsOCT further to detect even more numbers of different structures, we have synthesized axial submicron structure with layer thickness as 444.70 nm, 451.90 nm, 462.77 nm, 473.61 nm, 488.10 nm and with average index 1.383 ± 0.001 (**Fig. 2(a)** and **Fig. 2(c)**). Then, we have numerically simulated optical coherence tomography by illuminating broadband light with wavelength range from 1176 nm to 1413 nm. We have constructed nsOCT (**Fig. 2(b)** and **Fig. 2(d)**) following our developed methodology (see nsOCT flowchart in ESI: **Fig. S2**).

In another advance approach, we have numerically demonstrated that we can detect randomised submicron dominant synthetic structure at different depth with ~ 5 nm error (see **Fig. 2(e)** and **Fig. 2(f)**).

## Results and discussions

It is a general practice of large research community to test any new technique to apply in a system which is close to human and does not involved human sample study initially. Application a new technique on animal samples also gives opportunity to perform required numbers of experiment to optimize experimental protocols, post processing, image analysis etc. Therefore, we have decided to use mouse mammary fat pad (MFP) to test applicability of the comparably new nsOCT technique without sudden jumping towards human tissue characterization. Here efforts are to demonstrate new technical capabilities which proved can be deployed in future to access human skin via a hand-held probe or can be adopted with endoscopic system for imaging of intestine wall for early disease diagnosis based on nano-sensitive structural detection.

### Validation of nano-sensitive optical coherence tomography (nsOCT) numerically and experimentally in submicron axial structure.

In validation purpose, we have numerically constructed OCT images from different synthetic axial structures (axial structure size mentioned on each OCT B-scan, **Fig. 1(a)**). **Fig. 1(b)** displays numerically constructed nsOCT following nsOCT methodology (see nsOCT flowchart in ESI: **Fig. S2**) and detected submicron axial structure mentioned on each OCT B-scan. It is clearly demonstrated that our proposed numerical technique can detect axial submicron structure with ~ 1.5 nm accuracy (compare **Fig. 1(a)** and **Fig. 1(b)**). As a proof of concept experiment, state of art nsOCT approach applied on spectral OCT signal from two fabricated samples with two different repeating axial periodic submicron structures of 441.67 nm and 431.56 nm, respectively. Interference spectra obtained using spectral domain OCT (SDOCT) system (TELESTO III, Thorlabs, Inc.) and constructed OCT images (B-Scan) are presented in **Fig. 1(c)**. Subsequently, we have applied nsOCT approach on spectral OCT signal to extract dominant spatial period at each voxel of each Aline. These nsOCT images (B-Scan) were formed as colour maps of detected spatial periods and displays in **Fig. 1(b)** (numerically constructed) and **Fig. 1(d)** (experimentally). Hence, all areas with similar axial structure are visualized with the same colour. The depth of the preliminary nsOCT image was limited to ~ 1.0 mm by thresholding, and further optimization could increase the depth of access. Note that, in this experimental study, our detection accuracy is ~ 5 nm. Therefore, we can confirm that numerical simulation and experimental approach can detected and differentiate dominant nanostructures in different samples through state of art nano-sensitive optical coherence tomography (nsOCT) methodology with 5 nm accuracy. Although we have validated our approach numerically and experimentally, still there may have a question whether our proposed technique is applicable if we have different submicron structures at different depth? To check applicability of our proposed nsOCT method, we have synthesized five layers with thickness ~ 0.5 mm each and composed with five different submicron structures like 444.70 nm, 451.90 nm, 462.77 nm, 473.61 nm, 488.10 nm and placed in five different depth to construct a B-Scan. In **Fig. 2(a)**, we have displayed constructed B-Scan OCT image of these synthesized layers (structural composition of each layers mentioned on the B-Scan) and corresponding dominant spatial period (B-Scan) where we can see detected submicron structure with ~ 1.5 nm accuracy (compare **Fig. 2(a)** and **Fig. 2(b)**). Now, if we interchange each layer with their respective submicron structures throughout the depth, we can again detect them with ~ 1.5 nm accuracy (see results in **Fig. 2(c)** and **Fig. 2(d)**). Therefore, we have demonstrated that different dominant submicron structures at different depths can be detected through proposed state of art nano-sensitive optical coherence tomography (nsOCT) technique. However, further confirmation is required regarding applicability of nsOCT in practical situation, for example biological tissue samples composed of randomized submicron structure throughout the depth and does not expect to follow layer structure what we have demonstrated in **Fig. 2(a)-2(d)**. Therefore, in an advance numerical simulation approach (see **Fig. 2(e) and 2(f)**), we have demonstrated detection of randomised synthetic submicron structure. We have displays synthetic submicron structures (Y-Axis: sub-voxel axial structure) through the depth (X-Axis) in blue colour line. Here, synthetic submicron axial structures vary from 0.445 $\mu m$ to 0.488 $\mu m$. Subsequently, we have displayed detected axial sub-voxel dominant structures with $\sim 5$ nm accuracy in black colour plot throughout the depth using proposed nsOCT methodology. Eventually, this demonstration confirmed the applicability and reliability of our proposed nsOCT technique to detect depth resolved submicron structures with few nanometre accuracies.

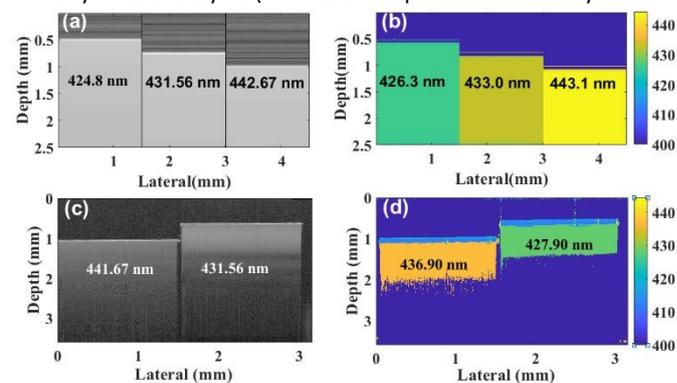

***Figure 1:*** *Simulated (a) OCT (present submicron scale structure mentioned) (b) nsOCT (detected submicron scale structure with colour bar), Experimental (c) OCT (present submicron scale structure mentioned), (d) nsOCT (colour bar represents detected submicron scale structure).*

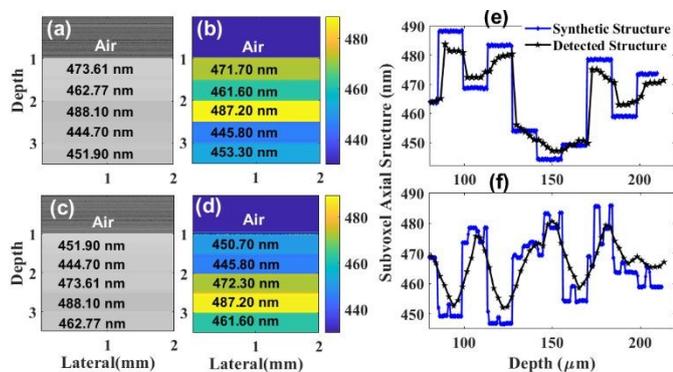

*Figure 2. (a) and (c) Represents OCT images with different submicron scale structure in different depth (simulated submicron scale structures mentioned at each layer). (b) and (d) represents maximum spatial period images with different submicron scale structure in different depth (detected sub-micron scale structures mentioned at each layer). Detection of randomised submicron structure with few nanometre accuracies. (e) Blue colour line (diamond) displays synthetic submicron structures throughout the depth. Black colour plot displayed detected axial sub-voxel dominant structures (maximum or most existing spatial period) throughout the depth using nsOCT methodology. (f) Blue colour line (diamond) displays synthetic submicron structures (synthetic size varies randomly over the depth) throughout the depth. Black colour plot displayed detected axial sub-voxel dominant structures (maximum spatial period) throughout the depth using nsOCT methodology.*

**Detection of sub-voxel informative nano-sensitive submicron structure in healthy and tumour tissue region from mouse**

After successful validation of the nsOCT approach via numerical simulation and experimental approach, we have applied our nsOCT techniques on fresh dissected tissue samples with healthy and tumour portion. **Fig. 3(a)** and **Fig. 3(b)** represents conventional H & E stained histological images from healthy and tumour tissue surface.

We can notice some qualitative structural difference between them but unable to quantify based on structural changes and it is also not a label free technique. **Fig. 3(c)** and **Fig. 3(d)** represents conventional OCT volume constructed from healthy and tumour tissue, respectively. Note that, it is impossible to differentiate from healthy to tumour tissue visually in OCT as it loses submicron structural information in conventional OCT construction (Axial resolution in tissue is $\sim 4.0$ μm, lateral resolution $\sim 13.0$ μm). Note that, resolution might be degraded over depth due to dispersion mismatch and other issue. However, we are not expecting resolution change have effect in our study much as nsOCT detection is based on spectral features of the interference signal and can detect submicron dominant structures which is much smaller than resolution itself.

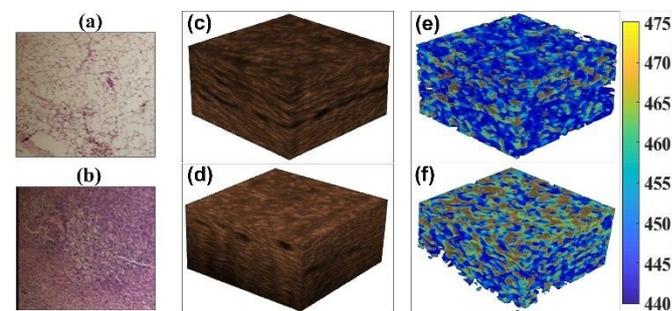

*Figure 3: H & E stained histological images (200 μm X 200 μm) of (a) normal and (b) tumour tissue. (c) and (d) 3D volume (250 μm X 250 μm X 180 μm) OCT images of healthy and tumour tissue. (e) and (f) Dominant submicron axial structural mapped 3D volume OCT images of healthy and tumour tissue, respectively. Colour bar represents size of dominant submicron axial structures.*

It is also worthy to mention here that lateral resolution simply can be improved by using high NA objective but will reduce the depth of access. Axial resolution also can be improved by using higher wavelength band OCT system. In this present research, the lateral resolution far larger than detected dominant axial structure and it is expected to have different dominant axial structure at different lateral (2D) location within resolution itself. In principle, we can detect different dominant axial structural information at different location in resolution by performing dense scanning and accruing different A-lines. In this study, we have also performed dense scan over 1.0 μm interval providing 13X13 different values of dominant structure within a lateral resolution. Note that, we have also axial resolution around 5.0 μm and even more in nsOCT images (~20.0 μm considering whole spectra divided in four windows for nsOCT construction) as we are doing windowing over interference spectra to construct nsOCT. Here, main theme is that we have able to detect submicron scale dominant structure within a 13 μm X 13 μm X 20 μm voxel. Regarding imaging data volume, it is essential to mention that nsOCT technique required very less amount of memory to process and store data compared to other nanoscale imaging technique with relatively same amount of information from same volume of interest. For example, with 10 nm resolution imaging technique, it required 13000 X 13000 X 20000 voxels to process and

Nanoscale Advances

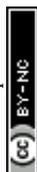

store images compared to single voxel in nsOCT with similar amount of information and with same volume of interest. **Fig. 3(e)** and **Fig. 3(f)** displays volume nsOCT where dominant axial submicron structure (most existing spatial period in a voxel or maximum spatial period) mapped onto healthy and tumour tissue, respectively. Notice that, these dominant nano structural maps provide clear differences between healthy and tumour tissue. Observation: lower submicron structure (0.445 μm – 0.455 μm) more prominent in healthy tissue where higher submicron structure (0.460 μm – 0.470 μm) is more prominent on tumour tissue. Details enface vs depth scan video with comparison between healthy and tumour submicron structure can be found in our electronic supplementary information (ESI) video files (ESI_video1: EnfaceHZvs Depth_Healthy.avi, ESI_video2: EnfaceHZvsDepth_Tumour.avi). Although, it is clear from enface images that there is an alteration in submicron dominant structure, it is essential to quantify overall structural distribution over enface images to differentiate tumour from healthy portion clinically. **Fig. 4(a)** and **Fig. 4(b)** displays histogram of dominant structure over superficial enface for healthy and tumour tissue region, respectively. **Fig. 4(c)** displays distribution of axial submicron structure throughout the volume of interest. Similar trends obvious (smaller and larger than 455 nm structure dominating in healthy and tumour tissue respectively) as observed in (**Fig. 4(a)** and **Fig. 4(b)**) axial structural distribution in first enface image from healthy and tumour tissue. We can clearly see alteration in dominant structural distribution from healthy to tumour: structure bellow 455 nm is more populated in healthy tissue in contrast to tumour tissue where axial structure larger than 455 nm is more populated. Note that, we have a limitation of our detection from 444 nm to 488 nm which is limited by wavelength range. We have found that 455 nm structure are more populated in tumour tissue compare to healthy one within our detection range (444 nm to 488 nm). There might be other axial structure which also getting populated in tumour tissue beyond our detection range and need to study further with an OCT system having higher wavelength band source. In this direction, our research group developing new ultra-high wavelength band OCT system for further investigation. For statistical significance and clinical applicability, it is advisable to look on statistics or size distribution at each depth (enface) and their alteration as tumour progress in multiple samples. In this direction, we required to perform nsOCT study on as many samples as possible to find consistent and statistically significant differences if any.

Volume (250 μm X 250 μm X 180 μm) histogram represents overall distribution of dominant submicron structure throughout healthy (blue colour) and tumour (red colour) tissue samples.

In supplementary **Fig. S3** (See **ESI**), we have displayed mean of dominant submicron structure over each enface corresponding to each depth of access. We have performed averaging of dominant structure over eight healthy tissue spots and ten tumour spots with area (250 μm X 250 μm) each. Higher trends of mean dominant submicron axial structure through analysed depth indicates increasing trends of size of submicron structures in tumour tissue compare to healthy one. Observations: overall ~10 nm changes occur close to tissue surface in transition from healthy to tumour. Possible reason may be due to swelling of cells, cell organelles and extra cellular vesicles in tumour progress.

## Conclusions

In conclusion, a Fourier domain nano-sensitive optical coherence tomography (nsOCT) method is validated, adapted, and explored for extraction and quantification of submicron scale axial structure with few nanometre accuracies. This adapted nsOCT technique successfully applied in depth wise randomised synthetic submicron axial structure with ~ 5 nm detection accuracy. Experimental validation of this nsOCT construction method on fabricated samples with different submicron structure and its detection with similar accuracy, and the agreement between the synthesized and nsOCT extracted structure, establishes self-consistency. We have also demonstrated that subtle changes (few nanometre) in depth resolved submicron structure can indeed be probed and quantified from OCT interference spectra, despite numerous complexities associated with the wide range of axial scaling at different depth. Subsequently, developed methodology applied in biological tissues to extract depth wise sub-voxel axial submicron structure. Results suggested that size of the overall submicron dominant structure increase by ~ 10 nm as tumour progress. The demonstrated ability to delineate submicron scale axial structural properties may provide a valuable non-invasive tool for characterization of tissue and wide variety of other complex scattering media of non-biological origin with few nanometre accuracies. Observed differences in the submicron scale axial morphology distribution in tissue volume between healthy and tumour shows considerable promise as a potential biomarker for early cancer detection and treatment response monitoring. These abilities to probe and quantify nano-sensitive morphological and structural alterations associated with pre-cancer using the nsOCT in backscattering mode bodes well for *in-vivo* deployment, because the backscattering geometry is clinically convenient, and the measurement technique is straight-forward. Exploiting the elastic backscattering spectra (singly or weakly scattered) recorded from tissue depths by interfering with reference signal, *in-vivo* applications of this approach can be realized with a suitably designed fibre optic probe. Finally, adapted nsOCT method represents as a fundamentally new and novel approach with much potential, but its actual benefits for *in-vivo* detection of precancers remain to be rigorously evaluated.

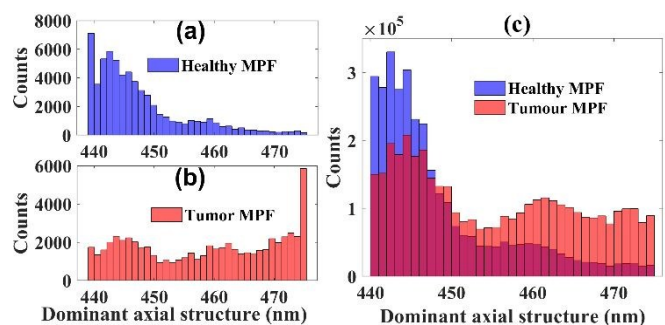

**Figure 4:** *(a) and (b) Histogram of dominant structure over superficial enface slice for healthy and tumour tissue regions, respectively. (c)*

## Conflicts of interest

There are no conflicts to declare.

## Acknowledgements

This project received funding from Irish Research Council (IRC), under Government of Ireland postdoctoral fellowship with project ID: GOIPD/2017/837. Nandan Das acknowledge National University of Ireland Galway (NUIG) for research facilities.

Also, this project has received funding from the European Union's Horizon 2020 research and innovation program under grant agreements no. 761214 and no. 779960. The materials presented and views expressed here are the responsibility of the author(s) only. The EU Commission takes no responsibility for any use made of the information set out.

This work was supported by NUI Galway, Galway University Foundation, the University of Limerick Foundation, the National Biophotonics Imaging Platform (NBIP) Ireland funded under the Higher Education Authority PRTLI Cycle 4 and co-funded by the Irish Government and the European Union.

## Ethical statements

All animal procedures were performed in accordance with the Guidelines for Care and Use of Laboratory Animals of "the Animal Care Research Ethics Committee (ACREC), National University of Ireland Galway (NUIG)" and approved by the "Health Product Regularity Authority (HPRA), Ireland".

## Notes and references

*a*Tissue Optics and Microcirculation Imaging (TOMI), National University of Ireland, Galway, Ireland
*b*Discipline of Surgery, Lambe Institute for Translational Research, National University of Ireland Galway, Ireland
*c*Institute of Photonic Sciences (ICFO), Barcelona, Spain
*d*Department of Biomedical Engineering (IMT), Linköping University, Sweden.
*Contact author: nandankds@gmail.com

Nanoscale Advances Accepted Manuscript

Please do not adjust margins

Nanoscale Advances

**ARTICLE**